\documentclass[12pt]{article}%
\usepackage[fleqn]{amsmath}
\usepackage{amssymb,euscript,theorem}%
\newcommand{\ncom}{\newcommand}%
\ncom{\rcom}{\renewcommand}%
\theoremstyle{plain}%
\theoremheaderfont{\normalfont\bfseries}%
\theorembodyfont{\itshape}%
\newtheorem{Lem}{Lemma}%
\newtheorem{Thm}{Theorem}%
\newtheorem{Corr}{Corollary}%
\ncom{\bskip}{\mbox{}\protect\\[0.5em]\noindent}%
\ncom{\half}{\mbox{\small $\frac{1}{2}$}}%
\rcom{\emptyset}{\varnothing}%
\ncom{\la}{\langle}%
\ncom{\ra}{\rangle}%
\ncom{\quadand}{\quad\mrm{and}\quad}%
\ncom{\quadiff}{\quad\mrm{iff}\quad}%
\ncom{\msf}{\mathsf}%
\ncom{\mbf}{\mathbf}%
\ncom{\tbf}{\textbf}%
\ncom{\trm}{\textrm}%
\ncom{\mrm}{\mathrm}%
\ncom{\bfm}{\boldmath}%
\ncom{\mcal}{\mathcal}%
\ncom{\mfrak}{\mathfrak}%
\ncom{\eul}{\EuScript}%
\ncom{\ensmath}{\ensuremath}%
\ncom{\terug}{\hspace*{-1.0ex}}%
\ncom{\wS}{\widehat{S}}%
\ncom{\wP}{\widehat{P}}%
\ncom{\wQ}{\widehat{Q}}%
\ncom{\wbfS}{\widehat{\bfS}}%
\ncom{\eulA}{\eul{A}}%
\ncom{\eulB}{\eul{B}}%
\ncom{\zero}{\textit{0}\,}%
\ncom{\one}{\textit{1}\,}%
\ncom{\calH}{\mcal{H}}%
\ncom{\calP}{\mcal{P}}%
\ncom{\calS}{\mcal{S}}%
\ncom{\calL}{\mcal{L}}%
\ncom{\calD}{\mcal{D}}%
\ncom{\bfe}{\mbf{e}}%
\ncom{\mbb}{\mathbb}%
\ncom{\perm}{\ensmath{\mbb{P}}}%
\ncom{\N}{\ensmath{\mbb{N}}}%
\ncom{\R}{\ensmath{\mbb{R}}}%
\ncom{\C}{\ensmath{\mbb{C}}}%
\ncom{\rmi}{\mrm{i}}%
\ncom{\sfC}{\msf{C}}%
\ncom{\sfP}{\msf{P}}%
\ncom{\sfT}{\msf{T}}%
\ncom{\longto}{\longrightarrow}%
\rcom{\to}{\rightarrow}%
\ncom{\mbbm}[1]{\mbox{\boldmath{$#1$}}}%
\ncom{\conj}{\mbbm{\;\wedge\;}}%
\ncom{\disj}{\mbbm{\;\vee\;}}%
\ncom{\deny}{\raisebox{0.2ex}{\boldmath{$\neg$}}}%
\ncom{\ifthen}{\mbbm{\;\longto\;}}%
\ncom{\desda}{\mbbm{\;\longleftrightarrow\;}}%
\ncom{\entails}{\mbbm{\vdash}}%
\ncom{\all}{\mbbm{\forall}\,}%
\ncom{\eris}{\mbbm{\exists}\,}%
\ncom{\Tr}{\mbox{\trm{Tr}}}%
\ncom{\tbfit}[1]{\textbf{\textit{#1}}}%
\ncom{\pa}{\tbfit{a}}%
\ncom{\pb}{\tbfit{b}}%
\ncom{\pc}{\tbfit{c}}%
\ncom{\pd}{\tbfit{d}}%
\ncom{\pone}{\tbfit{1}}%
\ncom{\ptwo}{\tbfit{2}}%
\ncom{\pthree}{\tbfit{3}}%
\ncom{\pj}{\tbfit{j}}%
\ncom{\pk}{\tbfit{k}}%
\ncom{\pN}{\tbfit{N}}%
\ncom{\fd}{\textsc{fd}}%
\ncom{\be}{\textsc{be}}%
\ncom{\mb}{\textsc{mb}}%
\ncom{\bfq}{\mbbm{q}}%
\ncom{\bfs}{\mbbm{s}}%
\ncom{\bfS}{\tbf{S}}%
\ncom{\fn}{\footnote}%
\ncom{\qm}{\mbox{\large \textsc{qm}}}%
\ncom{\qft}{\mbox{\large \textsc{qft}}}%
\ncom{\QM}{\trm{QM}}%
\ncom{\pii}{\mbox{\large \textsc{pii}}}%
\rcom{\it}{\mbox{\large \textsc{it}}}%
\ncom{\beq}{\begin{equation}}%
\ncom{\enq}{\end{equation}}%
\ncom{\meer}{\\*[0.5em]}%
\begin{document}%
\raggedbottom%
\thispagestyle{empty}%
\noindent
\begin{flushright}\fbox{To appear in: \textsc{philosophy of science}}
\end{flushright}\mbox{}\newline\newline
\begin{center}
\tbf{\LARGE Discerning Elementary Particles}\\[2em]
{\large F.A.\ Muller\fn{Faculty of Philosophy, Erasmus University Rotterdam,
Burg.\ Oudlaan 50, H5--16, 3062 PA Rotterdam,
E-mail: f.a.muller@fwb.eur.nl; and:
Institute for the History and Foundations of Science, Utrecht University,
Budapestlaan 6, IGG--3.08, 3584 CD Utrecht, The Netherlands
E-mail: f.a.muller@uu.nl} \& M.P.\ Seevinck\fn{Institute for the
History and Foundations of Science, Utrecht University,
Budapestlaan 6, IGG--3.08, 3584 CD Utrecht, The Netherlands
E-mail: m.p.seevinck@uu.nl}} \\[1em]
Utrecht, 30 April 2009 \\
Circa 7,000 words \\[1em] \mbox{}
\end{center}
\begin{quote}
\noindent{\small \tbf{Abstract.}~~We extend the
quantum-mechanical results of Muller \& Saunders (2008) establishing the
\emph{weak discernibility} of an arbitrary number of
similar fermions in finite-dimensional Hilbert-spaces in two ways:
(a) from fermions to bosons for all finite-dimensional Hilbert-spaces;
and (b) from finite-dimensional to infinite-dimensional Hilbert-spaces
for all elementary particles. In both cases this is performed using operators whose physical significance is beyond doubt.
This confutes the
currently dominant view that (A) the quantum-mechanical description of
similar particles conflicts with
Leibniz's Principle of the Identity of Indiscernibles ($\pii$);
and that (B) the only way to save $\pii$
is by adopting some pre-Kantian metaphysical notion such as Scotusian
haecceittas or Adamsian primitive thisness. We take sides with
Muller \& Saunders (2008) against this currently dominant view,
which has been expounded and defended by, among others,
Schr\"{o}dinger, Margenau, Cortes,
Dalla Chiara, Di Francia, Redhead, French, Teller, Butterfield, Mittelstaedt,
Giuntini, Castellani, Krause and Huggett.}
\end{quote}
\clearpage
\noindent\thispagestyle{empty}
\begin{quote}\tableofcontents\end{quote}
\thispagestyle{empty} \clearpage
\setcounter{page}{1}%
\setcounter{footnote}{0}%
\section{Introduction}
\label{SectIntro}
According to the founding father of wave mechanics Erwin Schr\"{o}dinger
(1996: 121--122), one of the ontological lessons that quantum mechanics ($\qm$)
has taught us is, as he told an audience in Dublin, February 1950,
that the elementary building blocks of the physical world are entirely
\emph{indiscernible}:
\begin{quote}{\small I beg to emphasize this and I beg you to believe it: it is not a
question of our being able to ascertain the identity in some instances and not being able
to do so in others. It is beyond doubt that the question of the `sameness', of identity,
really and truly has no meaning.}\end{quote}
Similar elementary particles have no `identity', there is nothing that discerns one particle from another,
neither properties nor relations can tell them apart, they are not individuals.
Thus Schr\"{o}dinger famously compared the elementary particles to ``the shillings and pennies in your bank account'', in contrast to the coins in your piggy-bank. In 1928,
Hermann Weyl (1950: 241) had preceded Schr\"{o}dinger when he wrote
that ``even in principle one cannot demand an alibi from an electron''.

Over the past decades, several philosophers have scrutinised
this \emph{Indiscernibility Thesis} ($\it$)
by providing various rigorous arguments in favour
of it: \emph{similar} elementary particles (same mass, charge, spin, etc.),
when forming a composite physical system, are indiscernible by quantum-mechanical means.
Leibniz's metaphysical Principle of the Identity of Indiscernibles ($\pii$)
is thus refuted by physics ($\qm$) --- and perhaps is therefore not so metaphysical after all. This does not rule out conclusively that particles
\emph{really} are discernible,
but \emph{if} they are, they have to be discerned by means that go above and beyond physics ($\qm$), such as by ascribing Scotusian \emph{haecceitas} to the particles, or
ascribing sibling attributes to them from scholastic and
neo-scholastic metaphysics. Nevertheless
few philosophers have considered this move to save
the discernibility of the elementary particles to be attractive ---
if this move is mentioned, then usually as a possibility and rarely
as a plausibility. Mild naturalistic inclinations seem sufficient to
accommodate $\it$ in our general metaphysical view of the
world. We ought to let well-established scientific knowledge
inform our metaphysical view of the world whenever possible and appropriate,
and this is exactly what Schr\"{o}dinger begged us to do. The only
respectable metaphysics is naturalised metaphysics; see further
Ladyman \& Ross (2007: 1--38). Prominent defenders of $\it$ include:
Margenau (1944); Cortes (1976), who brandished $\pii$
``a false principle''; Barnette (1978), French
\& Redhead (1988); Giuntini \& Mittelstaedt (1989),
who argued that although demonstrably valid in classical logic, in
quantum logic the validity of $\pii$ cannot be established;
French (1989a), who assured us in the title that $\pii$ ``is
not contingently true either''; French (1989a; 1989b; 1998; 2006),
Redhead \& Teller (1992), Butterfield (1993), Castellani \& Mittelstaedt (2000),
Massimi (2001), Teller (1998), French \& Rickles (2003), Huggett (2003),
French \& Krause (2006: Ch.$\,4$).

There have however been dissenters. B.C.\ van Fraassen (1991)
is one of them; see Muller \& Saunders (2006: 517--518) for an
analysis of his arguments. We follow the other dissenters: S.W.\
Saunders and one of us (Saunders (2006), Muller \& Saunders (2008)).
They neither claim that fermions are individuals nor do they rely on a
particular interpretation of $\qm$. On the basis of standard mathematics
(standard set-theory and classical predicate logic) and only uncontroversial
postulates of $\qm$ (notably
leaving out the projection postulate, the strong property postulate and
the quantum-mechanical probabilities),
they demonstrate that similar \emph{fermions} are \emph{weakly discernible}, i.e.\
they are discerned by relations that are irreflexive and symmetric,
in every admissible state of the composite system. So according to Muller \& Saunders,
the elementary building blocks of matter (fermions) are \emph{not}
indiscernibles after all, contra $\it$. They prove this, however, only for
\emph{finite-dimensional} Hilbert-spaces (their Theorem$\,1$),
which is a rather serious restriction because most applications of $\qm$ to physical systems employ complex wave-functions and these live in the \emph{infinite-dimensional} Hilbert-space $L^{2}(\R^{3N})$; nonetheless they confidently
conjecture that their result will hold good for infinite-dimensional Hilbert-spaces
as well (their Conjecture$\,1$).
Furthermore, Muller \& Saunders (2008: 534--535) need to \emph{assume} for their proof
there is a maximal self-adjoint operator acting on finite-dimensional
Hilbert-spaces that is physically significant.
In the case of dimension $2$ of a single-fermion Hilbert-space, Pauli's spin-$1/2$
operator qualifies as such a maximal self-adjoint operator, but for higher dimensions, spin is degenerate. What this maximal operator in those cases corresponds
to they gloss over.

When it comes to the elementary
quanta of interaction (bosons), Muller \& Saunders (2008: 537--540) claim that
bosons are also weak discernibles, but of a \emph{probabilistic} kind
(the discerning relation involves quantum-mechanical probabilities and therefore their
proof needs the Probability Postulate of $\qm$),
whereas the discerning relation of the fermions is of a \emph{categorical}
kind (no probabilities involved). More precisely, the categorical discernibility of bosons turns out to be a contingent matter: in some states they are categorically discernible,
in others, e.g.\ direct-product states, they are not; this prevents one
to conclude that bosons are categorical discernibles \emph{simpliciter}.
But the boson's probabilistic discernibility is a quantum-mechanical necessity;
Theorem$\,3$ (\emph{ibid.}) establishes it for two bosons, with no restrictions on the dimensionality of Hilbert-space but conditional on whether a particular sort of operator can be found (again, a maximal self-adjoint one of physical
significance). The fermions are also probabilistic discernibles; their Theorem$\,2$ states it for finite-dimensional Hilbert-spaces only and is therefore equally restrictive
as their Theorem$\,1$.

The central aim of the current paper is the completion of the project initiated and
developed in Muller \& Saunders (2008), by demonstrating that
\emph{all} restrictions in \emph{their} discernibility theorems can be removed by proving
more general theorems and proving them differently than they have done, employing only quantum mechanical operators that have obvious physical significance. We shall then be in a position to conclude in utter
generality that \emph{all} kinds of similar particles in \emph{all} their
physical states, pure and mixed, in \emph{all}
infinite-dimensional or finite-dimensional Hilbert-spaces can be \emph{categorically}
discerned on the basis of quantum-mechanical postulates.
This result, then, should be the death-knell for $\it$, and, by implication,
establishes the universal reign of Leibniz's $\pii$ in $\qm$.

In Sections$\,$\ref{SectDiscThmsInf} and \ref{SectDiscThmsFin},
we prove the theorems that establish the general result.
First we introduce some
terminology, state explicitly what we need of $\qm$,
and address the issue of what has the license to discern
elementary particles (Section$\,$\ref{SectPrelim}).
\section{Preliminaries}
\label{SectPrelim}
For the motivation and further elaboration of the terminology
we are about to introduce, we refer to Muller \& Saunders
(2008: 503--505) because we follow them closely
(readers of that paper can jump to the next Section of the
current paper). Here we only mention what is necessary
in order to keep the current paper comparatively self-contained.

We call physical objects in a set \emph{absolutely discernible},
or \emph{individuals}, iff for every
object there is some physical property that it has but all
others lack; and \emph{relationally discernible} iff for every
object there is some physical relation that discerns it from all
others (see below). An object is \emph{indiscernible} iff it is both absolutely and
relationally indiscernible, and hence \emph{discernible} iff it is
discernible either way or both ways. Objects that are not
individuals but are relationally discernible from all other objects we call
\emph{relationals}; then \emph{indiscernibles} are objects that are neither
individuals nor relationals.
Quine (1981: 129--133) was the first to inquire into
different kinds of discernibility; he discovered there are
\emph{only two} independent
logical categories of relational discernibility (by means of a
binary relation): either the relation is irreflexive and
asymmetric, in which case we speak of \emph{relative
discernibility}; or the relation is irreflexive and symmetric, in
which case we speak of \emph{weak discernibility}. We call attention
to the logical fact that
if relation $R$ discerns particles $\pone$ and $\ptwo$
\emph{relatively}, then its \emph{complement relation}, defined as
$\deny R$, is also asymmetric but reflexive; and if $R$
discerns particles $\pone$ and $\ptwo$ \emph{weakly}, then its
complement relation $\deny R$ is reflexive and symmetric but does
not hold for $\pa\neq\pb$ whenever $R$ holds for $\pa\neq\pb$.

Leibniz's \emph{Principle of the Identity of Indiscernibles}
($\pii$) for physical objects states that \emph{no two
physical objects are absolutely and
relationally indiscernible; or synonymously, two physical
objects are numerically discernible only if they are
qualitatively discernible.}
One can further distinguish principles for absolute
and for relational indiscernibles and then inquire into the
logical relations between these and $\pii$; see
Muller \& Saunders (2008: 504--505).
Similarly one can distinguish three indiscernibility
theses as the corresponding negations of the Leibnizian principles.
We restrict ourselves to the \emph{Indiscernibility Thesis} ($\it$):
\emph{there are composite
systems of similar physical objects that consist of absolutely and
relationally indiscernible physical objects.}
Then either it is a theorem of logic that $\pii$
holds and $\it$ fails, or conversely:
\beq \entails\;\;\pii\;\desda\;\deny\it\;. \label{PIIIT}\enq

Next we rehearse the postulates of $\qm$ that we shall use
in our Discernibility Theorems.

The \emph{State Postulate} (StateP) associates some super-selected
sector Hilbert-space $\calH$ to every given
physical system $S$ and represents every \emph{physical state} of $S$ by
a statistical operator $W\in\calS(\calH)$; the pure states lie on the boundary
of this convex set $\calS(\calH)$ of all statistical operators and the mixed states
lie inside. If $S$ consists of $N$ similar elementary particles, then the
associated Hilbert-space is a direct-product Hilbert-space
$\calH^{N}=\calH\otimes\cdots\otimes\calH$ of $N$ identical
single-particles Hilbert-spaces.

The \emph{Weak Magnitude Postulate} (WkMP) says that every physical magnitude is represented by an operator that acts on $\calH$. Stronger magnitude postulates
are not needed, because they all imply the logically weaker WkMP,
which is sufficient for our purposes.

In order to state the Symmetrisation Postulate, we need to define
first the orthogonal projectors $\Pi^{\pm}_{N}$ of the
lattice $\calP(\calH^{N})$ of all projectors, defined as
\beq \Pi^{+}_{N}\equiv\frac{1}{N!}\sum^{N!}_{\pi\in\perm_{N}}U_{\pi}
\;\quadand\;\Pi^{-}_{N}\equiv\frac{1}{N!}
\sum^{N!}_{\pi\in\perm_{N}} \trm{sign}(\pi)U_{\pi}\;,
\label{DefPi} \enq
where sign$(\pi)\in\{\pm 1\}$ is the \emph{sign} of the permutation
$\pi\in\perm_{N}$ on $\{1,2,\ldots,N\}$ ($+1$ if it is even, $-1$ if odd),
and where $U_{\pi}$ is a unitary operator acting on $\calH^{N}$
corresponding to permutation $\pi$ (these $U_{\pi}$ form a
unitary representation on $\calH^{N}$ of the permutation group $\perm_{N}$).
The projectors \eqref{DefPi} lead to the
following permutation-invariant orthogonal subspaces:
\beq \calH_{+}^{N}\equiv\Pi^{+}_{N}\big[\calH^{N}\big]\;\quadand\;
\calH_{-}^{N}\equiv\Pi^{-}_{N}\big[\calH^{N}\big]\;, \label{HSHA} \enq
which are called the {\be}-\emph{symmetric} (Bose-Einstein)
and the {\fd}-\emph{symmetric} (Fermi-Dirac)
subspaces of $\calH^{N}$, respectively.
These subspaces can, alternatively, be seen as generated
by the symmetrised and anti-symmetrised versions of the
products of basis-vectors in $\calH^{N}$. Only for $N=2$,
we have that $\calH^{N}_{-}\oplus\calH^{N}_{+}=\calH^{N}$.

The \emph{Symmetrisation Postulate} (SymP) states
for a composite system of $N\geqslant 2$ similar particles with
$N$-fold direct-product Hilbert-space
$\calH^{N}$ the following: (i) the projectors $\Pi^{\pm}_{N}$ \eqref{DefPi}
are super-selection operators; (ii) integer and
half-integer spin particles are confined to the
{\be}- and the {\fd}-symmetric subspaces \eqref{HSHA}, respectively;
and (iii) all composite systems of similar particles consist
of particles that have all either integer spin or half-integer spin (Dichotomy).

We represent a \emph{quantitative physical property} associated with
physical magnitude $\eulA$ mathematically by ordered pair
$\la A,a\ra$, where $A$ is
the operator representing $\eulA$ and $a\in\C$. The \emph{Weak Property Postulate}
(WkPP) says that if
the physical state of physical system $S$ is an eigenstate of $A$
having eigenvalue $a$, then it has property $\la A,a\ra$; the
\emph{Strong Property Postulate} (StrPP) adds the converse
conditional to WkPP. (We mention that
eigenstates can be \emph{mixed}, so that physical systems
in \emph{mixed states} can posses properties too (by WkPP).
see Muller \& Saunders (2008: 513) for
details.) WkPP implies that every physical
system $S$ always has the same
quantitative properties associated with all super-selected
physical magnitudes because $S$ always is in the same common
eigenstate of the super-selected operators.
We call these
possessed quantitative physical properties \emph{super-selected}
and we call physical systems, e.g.\ particles, that have the same
super-selected quantitative physical properties \emph{similar} (this
is the precise definition of `similar', a word that we have
been using loosely until now, following Dirac).

We also adopt the following \emph{Semantic Condition} (SemC).
When talking of a physical system at a given time, we ascribe to it at most
\emph{one} quantitative physical property associated with physical
magnitude $\eulA$:
\beq \trm{(SemC)}~~\textit{If physical system $S$ possesses $\la A,a\ra$ and $\la A,a'\ra$,
then $\;a=a'\,$.} \label{SemC} \enq
For example, particles cannot possess two different masses at the same time and
\eqref{SemC} is the generalisation of this in the language of
$\qm$. In other words: if $S$ possesses quantitative physical
property $\la A,a\ra$, then $S$ does \emph{not} posses property
$\la A,a'\ra$ for every $a'\neq a$. Statement \eqref{SemC} is
neither a tautology nor a theorem of logic, but we agree with Muller \&
Saunders (2008: 515) in that ``it seems absurd to
deny it all the same''.

Notice there is neither mention of measurements
nor of probabilities in the postulates mentioned above,
let alone interpretational glosses such as dispositions.

For an outline of \emph{the elementary language of} $\qm$,
we refer again to Muller \& Saunders (2008: 520--521).
In this language, the proper formulation of $\pii$ is that
\emph{physically indiscernible physical systems are identical}
(Muller \& Saunders 2008: 521--523):
\beq \msf{PhysInd}(\pa,\;\pb)\;\;\ifthen\;\;\pa=\pb\;, \label{PIIQM}
\enq
where `$\pa$' and `$\pb$' are physical-system variables, ranging over all
physical systems,
and where $\msf{PhysInd}(\pa,\pb)$ comprises everything that
is in principle permitted to discern the particles: roughly,
all physical relations and all physical properties.
The properties and the relations may involve, in their
definition, probabilities, in which case we call them \emph{probabilistic};
otherwise, in the absence of probabilities, we call them \emph{categorical}.
So the three logical kinds of discernibility --- (a) absolute and (r) relational,
which further branches in (r.w) weak and (r.r) relative discernibility --- come
in a probabilistic and a categorical variety. In their analysis of the traditional
arguments in favour of $\it$, Muller \& Saunders (2008: 524--526)
make the case that, setting conditional probabilities aside, (r) relational
discernibility has been largely
overlooked by the tradition. (Parenthetically, Leibniz also included relations
in his $\pii$ \emph{because} he held that all relations reduce to properties
and thus could make do with an explicit formulation of $\pii$ that only
mentions properties; give up his reducibility thesis of relations to
properties and one can no longer make do with his formulation; see
Muller \& Saunders (2008: 504--505).) What in particular has been
overlooked, and is employed by Muller \& Saunders, are properties of wholes
that are relations between their constitutive parts: the distance between the Sun
and the Earth is a property of the solar system; the Coulomb-interaction
between the electron and the proton is a property of the Hydrogen atom; etc.

Muller \& Saunders (2008: 524--528) argue at length that only those
properties and relations are permitted to occur in $\msf{PhysInd}$ \eqref{PIIQM}
that meet the following two requirements.
\begin{itemize}
{\item[](Req1)~~\emph{Physical meaning}. All properties and relations,
as they occur WkPP, should be transparently defined in terms of physical
states and operators
that correspond to physical magnitudes in order for the
properties and relations to be physically meaningful.}
{\item[](Req2)~~\emph{Permutation invariance}. Any property of one particle is
a property of any other; relations should be permutation-invariant,
so \emph{binary} relations should be symmetric \emph{and} either reflexive or
irreflexive.}\label{Req12}
\end{itemize}

All proponents of the Indiscernibility Thesis ($\it$)
have considered quantum-mechanical means of discerning
similar particles that obey these two Requirements (see the references listed
in the Introduction) --- and have found them all to fail.
They were correct in this. They were not correct in \emph{not} considering
categorical relations.

To close this Section, we want address another distinction from the recent
flourishing literature on indiscernibility
and inquire briefly whether this motivates a \emph{third} requirement,
call it Req3. One easily shows that absolute discernibles are
relational discernibles by \emph{defining} a relation (expressed by
dyadic predicate $R_{M}$) in terms of
the discerning properties (expressed by monadic predicate $M$);
see Muller \& Saunders (2008: 529).
One could submit that this is not a case of `genuine' but of `fake' relational
discernibility, because there is nothing inherently relational about
the way this relational discernibility is achieved: $R_{M}$ is completely
reducible to property $M$, which already discerns the particles
absolutely. Similarly, one may also object
that a case of absolute discernibility implied by relational
discernibility by means of
a monadic predicate $M_{R}$ that is defined in terms of the
discerning relation $R$ is not a case of `genuine' but of `fake' absolute discernibility
(the terminology of `genuine' and `fake' is not Ladyman's (2007: 36),
who calls `fake' and `genuine' more neutrally ``contextual'' and ``intrinsic'',
respectively). Definitions: physical systems $\pa$ and $\pb$ are
\emph{genuine} relationals, or \emph{genuine} (weak, relative) relational discernibles,
iff they are discerned by some dyadic predicate \emph{that is not reducible to
monadic predicates of which some discern $\pa$ and $\pb$ absolutely};
$\pa$ and $\pb$ are \emph{genuine} individuals, or \emph{genuine}
absolute discernibles, iff they are discerned by some monadic predicate
\emph{that is not reducible to dyadic predicates
of which some discern $\pa$ and $\pb$ relationally}; discernibles are \emph{fake}
iff they are not genuine. Hence there is a \emph{prima facie} case for
adding a third Requirement that excludes fake discernibility:
\begin{itemize}
{\item[](Req3) \emph{Authenticity.} Predicates expressing
discerning relations and properties must be genuine.}
\end{itemize}

In turn, a \emph{fake} property or relation can be defined rigorously as its \emph{undefinability in terms
of the predicates in the language of $\qm$ that meet} Req1 \emph{and} Req2.
In order to inquire logically into genuineness and fakeness, thus
defined, at an appreciable level of rigour,
the entire formal language must be spelled out and all axioms of $\qm$ must be spelled out in that
formal language. Such a logical inquiry is however far beyond the scope
of the current paper. Nonetheless we shall see that our discerning relations plausibly \emph{are}
genuine.

But besides formalise-fobia, there is a respectable reason for not adding Req3
to our list. To see why, consider the following two cases: ($a$) indiscernibles
and ($b$) discernibles.

($a$) Suppose particles turn out to be indiscernibles in that they
are indiscernible by all genuine relations and all genuine properties.
Then they are also indiscernible by \emph{all} properties and \emph{all} relations
that are defined in terms of these, which one can presumably prove by
induction over the complexity of the defined
predicates. So indiscernibles remain indiscernibles, whether we require the candidate
properties and relations to be genuine or not.

($b$) Suppose next that the particles turn out to be discernibles.
($b.i$) If they are discerned by a relation that turns out to be definable
in terms of genuine properties one of which discerns the particles
absolutely, then the relationals
become individuals --- good news for admirers of $\pii$. But the important point to
notice is that discernibles remain discernibles. ($b.ii$) If the particles are discerned
by a property that turns turns out to be definable in terms of genuine relations
one of which discerns the particles relationally, then the individuals loose their
individuality and become relationals. They had a fake-identity and are
now exposed as metaphysical imposters.
But again, the important point to notice is that discernibles remain discernibles.

To conclude, adding Req3 will not have any consequences for crossing the border
between discernibles and indiscernibles. This seems a respectable reason not to add
Req3 to our list of two Requirements.
\section{To Discern in Infinite-Dimensional Hilbert-Spaces}%
\label{SectDiscThmsInf}
We first prove a Lemma, from which our categorical discernibility theorems then
immediately follow.
\begin{Lem}[\textbf{StateP, WkMP, WkPP, SemC}]~~Given a composite physical system of
$N\geqslant 2$ similar particles and its associated direct-product Hilbert-space
$\calH^{N}$. If there are two single-particle operators, $A$ and $B$,
acting in single-particle Hilbert-space $\calH$, and they correspond to physical
magnitudes $\eulA$ and $\eulB$, respectively,
and there is a non-zero number $c\in\C$ such that in every pure state
$|\phi\ra\in\calH$ in the domain of their commutator the following holds:
\beq [A,\,B]|\phi\ra\,=\,c|\phi\ra\;, \label{ABc} \enq
then all particles are categorically weakly discernible.
\label{Lem1}\end{Lem}
\noindent\tbf{Proof.}
Let $\pa,\pb,\pj$ be particle-variables, ranging over the set  $\{\pone,\ptwo,\ldots,\pN\}$ of $N$ particles.
We proceed Step-wise, as follows.\bskip
\indent [S1]~~Case for $N=2$, pure states.\newline
\indent [S2]~~Case for $N=2$, mixed states.\newline
\indent [S3]~~Case for $N>2$, pure states.\newline
\indent [S4]~~Case for $N>2$, mixed states.\bskip
[S1].~~\emph{Case for $N=2$, pure states.} Assume the antecedent. Define the following operators on $\calH^{2}=\calH\otimes\calH$:
\beq A_{\!\pone}\,\equiv\,A\otimes\one\quadand
A_{\!\ptwo}\,\equiv\,\one\otimes A\;, \label{DefA12} \enq
where the operator $\one$ is the identity-operator on $\calH$;
and \emph{mutatis mutandis} for $B$.
Define next the following \emph{commutator-relation}:
\beq \sfC(\pa,\pb)\quadiff\all |\Psi\ra\in\calD:\;
\big[A_{\pa},\,B_{\pb}\big]|\Psi\ra\,=\,c|\Psi\ra\;,
\label{DefC} \enq
where $\calD\subseteq\calH\otimes\calH$ is the domain of the commutator.
An arbitrary vector $|\Psi\ra$ can be expanded:
\beq |\Psi\ra=\sum_{j,k=1}^{d}\gamma_{jk}|\phi_{j}\ra\otimes|\phi_{k}\ra\;,
\qquad\sum_{j,k=1}^{d}|\gamma_{jk}|^2=1,
\label{ExpPsi} \enq
where $d$ is a positive integer or $\infty$, and $\{|\phi_{1},|\phi_{2}\ra,\ldots\}$
is a basis for $\calH$ that lies in the domain of the commutator $[A,B]$.
Then using expansion \eqref{ExpPsi} and eq.$\,$\eqref{ABc} one quickly shows
that ($\pa=\pb$):
\beq \label{commaa}
\big[A_{\pa},\,B_{\pa}\big]|\Psi\ra\,=\,c|\Psi\ra\;,
\enq
and that for $\pa\neq\pb$:
\beq\label{commab}
\big[A_{\pa},\,B_{\pb}\big]|\Psi\ra\,=\,0|\Psi\ra\,\neq\,c|\Psi\ra\;,
\enq
because by assumption $c\neq 0$. By WkPP, the composite system then possesses
the following \emph{four} quantitative physical properties (when substituting $\pone$ or
$\ptwo$ for $\pa$ in the first, and $\pone$ for $\pa$ and $\ptwo$ for $\pb$, or
conversely, in the second):
\beq \big\la\big[A_{\pa},\,B_{\pa}\big],\,c\big\ra\;
\quadand \big\la\big[A_{\pa},\,B_{\pb}\big],\,0\big\ra
\quad(\pa\neq\pb)\;. \label{Aux1}\enq
In virtue of the Semantic Conditional \eqref{SemC}, the composite system then
does not possess the following \emph{four} quantitative physical properties (recall
that $c\neq 0$):
\beq \big\la\big[A_{\pa},\,B_{\pa}\big],\,0\big\ra\;
\quadand\big\la\big[A_{\pa},\,B_{\pb}\big],\,c\big\ra
\quad(\pa\neq\pb)\;. \label{Aux2}\enq
The composite system possesses the property \eqref{Aux1} that
\emph{is} a relation between its constituent
parts, namely $\sfC$ \eqref{DefC}, which is reflexive:
$\sfC(\pa,\pa)$ for every $\pa$ due to \eqref{commaa}.
Similarly, but now using SemC \eqref{SemC},
the composite system does \emph{not} possess the
property \eqref{Aux2} that \emph{is} a relation between
its constituent parts, namely $\sfC$.
Therefore $\pone$ is not
related to $\ptwo$, and $\ptwo$ is not related to $\pone$ either, because
$\deny\sfC(\pa,\pb)$ and $\deny\sfC(\pb,\pa)$ ($\pa\neq\pb$);
and then, due to the following theorem of logic:
\beq \entails\;\;\big(\deny\sfC(\pa,\pb)\conj\deny\sfC(\pb,\pa)\big)\;\ifthen\;
\big(\sfC(\pa,\pb)\desda\sfC(\pb,\pa)\big)\;, \label{theoremlogic}\enq
we conclude that $\sfC$ is symmetric (Req2). Since by assumption
$A$ and $B$ correspond to
physical magnitudes, relation $\sfC$ \eqref{DefC}
is physically meaningful (Req1) and hence is admissible, because
it meets Req1 and Req2.

Further, it was just shown that the relation $\sfC$ \eqref{DefC} is
reflexive and symmetric but fails for $\pa\neq\pb$ due to \eqref{commab},
which means that $\sfC$
discerns the two particles \emph{weakly} in every
pure state $|\Psi\ra$ of the composite
system. Since probabilities do not occur
in $\sfC$ \eqref{DefC}, the particles are discerned \emph{categorically}.

[S2].~~\emph{Case for $N=2$, mixed states.}
The equations in \eqref{DefC} can also be written as an equation
for $1$-dimensional projectors that project onto the ray that contains
$|\Psi\ra$:
\beq \big[A_{\pa},\,B_{\pb}\big]|\Psi\ra\la\Psi|\,=\,c|\Psi\ra\la\Psi|\;.
\label{DefCPr} \enq
Due to the linearity of the operators, this equation remains
valid for arbitrary linear combinations of projectors.
This includes all convex combinations of projectors, which exhausts the set
$\calS(\calH\otimes\calH)$ of \emph{all} mixed
states. The commutator-relation $\sfC$ \eqref{DefC} is easily extended to mixed
states $W\in\calS(\calH\otimes\calH)$ and the ensuing relation also discerns
the particles categorically and weakly.

[S3], [S4].~~\emph{Case for $N>2$, pure and mixed states.}
Cases [S1] and [S2] are immediately extended to the $N$-particle cases, by considering
the following $N$-factor operators:
\beq A_{\!\pj}\,\equiv\,\one\otimes\;\cdots\;\otimes\one\otimes
A\otimes\one\otimes\;\cdots\;\otimes\one\;, \label{DefAj} \enq
where $A$ is the $j$-th factor and $\pj$ a particle-variable
running over the $N$ labeled particles, and similarly for $B^{(\pj)}$.
The extension to the mixed states then proceeds as in [S2]. \tbf{Q.e.d.}
\begin{Thm}[\textbf{StateP, WkMP, WkPP, SemC}]~~In a composite physical system of a finite number of
similar particles, all particles are categorically weakly
discernible in every physical state, pure and mixed,
for every infinite-dimensional Hilbert-space. \label{ThmDiscNPartInf}
\end{Thm}
\tbf{Proof.} In Lemma~\ref{Lem1}, choose for $A$ the linear momentum operator $\wP$, for $B$ the Cartesian position-operator $\wQ$, and for $c$ the value $-\rmi\hslash$.
The physical significance of these operators and their commutator, which
is the celebrated canonical commutator
\beq \big[\wP,\,\wQ\,\big]=-\rmi\hslash\one\;, \label{CCR} \enq
is beyond doubt and so is the ensuing commutator relation $\sfC$
\eqref{DefC}, which we baptise \emph{the Heisenberg-relation}.
The operators $\wP$ and $\wQ$ act on the infinite-dimensional Hilbert-space
of the complex wave-functions $L^{2}(\R^{3})$, which is isomorphic to \emph{every infinite-dimensional} Hilbert-space. \tbf{Q.e.d.} \bskip
\indent But is Theorem$\,$\ref{ThmDiscNPartInf} not only applicable to particles having
spin-$0$ and have we forgotten to mention this? Yes and No. Yes, we have
deliberately forgotten to mention this. No, it is a corollary of
Theorem$\,$\ref{ThmDiscNPartInf} that it holds for all spin-magnitudes,
which is the content of the next theorem.
\begin{Corr}[\textbf{StateP, WkMP, WkPP, SemC}]~~In a composite physical system
of $N\geqslant 2$ similar particles of arbitrary spin, all particles are categorically weakly discernible in every admissible physical state, pure and mixed,
for every infinite-dimensional Hilbert-space. \label{CorrSpin}
\end{Corr}
\tbf{Proof.} To deal with spin, we need SymP.
The actual proof of the categorical weak discernibility for
all particles having non-zero spin-magnitude is
at bottom a notational variant of the proof of
Theorem$\,$\ref{ThmDiscNPartInf}. Let us sketch how this works for $N=2$.
We begin with the following Hilbert-space for a single particle:
\beq \calH_{s}\,\equiv\,\big(L^{2}(\R^{3})\big)^{2s+1}\;, \label{Hs}
\enq
which is the space of \emph{spinorial wave-functions} $\Psi$, i.e.\ column-vectors
of $2s+1$-entries, each entry being a complex wave-function of $L^{2}(\R^{3})$.
Notice that $\calH_{s}$ is an $(2s+1)$-fold Cartesian-product set, which
becomes a Hilbert-space by carrying the Hilbert-space properties of $L^{2}(\R^{3})$
over to $\calH_{s}$. For instance, the inner-product on $\calH_{s}$ is
just the sum of the inner-products of the components of the spinors:
\beq \la\Psi|\Phi\ra\,=\,\sum_{k=1}^{2s+1}\la\Psi_{k}|\Phi_{k}\ra\;,
\enq
where $\Psi_{k}$ is the $k$-th entry of $\Psi$ (and 
similarly for $\Phi_{k}$), 
which provides the norm of $\calH_{s}$, which in turn generates the
norm-topology of $\calH_{s}$, etc.
The degenerate case of the spinor having only one entry is
the case of $s=0$, which we treated in Theorem$\,$\ref{ThmDiscNPartInf}.
So we proceed here with $s>0$. In particular, the number of
entries $2s+1$ is even iff the particles have half-integer spin,
and odd iff the particles have integer spin, in units of $\hslash$.

Let $\bfe_{k}\in\C^{2s+1}$ be such that its $k$-th
entry is $1$ and all others $0$. They form the standard
basis for $\C^{2s+1}$ and are the
eigenvectors of the $z$-component of the spin-operator
$\widehat{S}_{z}$, whose eigenvalues are traditionally denoted by $m$
(in the terminology of atomic physics: `magnetic quantum number'):
\beq \widehat{S}_{z}\bfe_{k}\,=\,m_{k}\bfe_{k}\;,
\enq
where $m_{1}=-s$, $m_{2}=-s+1$, $\ldots$, $m_{s-1}=s-1$, $m_{s}=+s$.
Let $\phi_{1},\phi_{2},\ldots$ be a basis for $L^{2}(\R^{3})$;
then this is a basis for single-particle spinor space $\calH_{s}$
\eqref{Hs}:
\beq \big\{\bfe_{k}\phi_{m}\in\calH_{s}\mid k\in\{1,2,\ldots,2s+1\},\;
m\in\N^{+}\big\}\;. \label{basisHs} \enq
Recall that the linear momentum-operator $\wP$ and the Cartesian position
operator $\wQ$ on an arbitrary complex wave-function $\phi\in L^{2}(\R^{3})$
are the differential-operator times $-\rmi\hslash$ and the
multiplication-operator, respectively:
\beq
\wP:\calD_{P}\to L^{2}(\R^{3}),\;
\phi\mapsto\wP\phi,\quad\trm{where}\;\;
\big(\wP\phi\big)(\bfq)\,\equiv\,
-\rmi\hslash\frac{\partial\phi(\bfq)}{\partial\bfq}
\label{DefP}  \enq
and
\beq \wQ:\calD_{Q}\to L^{2}(\R^{3}),\;
\phi\mapsto\wQ\phi,\quad\trm{where}\;\;
\big(\wQ\phi\big)(\bfq)\,\equiv\,(q_{x}+q_{y}+q_{z})\phi(\bfq)\;,
\label{DefQ}\enq
where domain $\calD_{P}=C^{1}(\R^{3})\cap L^{2}(\R^{3})$
and
domain $\calD_{Q}\subset L^{2}(\R^{3})$ consist of all wave-functions
$\psi$ such that $|\bfq|^{2}\psi(\bfq)\to 0$
 when $|\bfq|\to\infty$.
The action of $\wP$ and $\wQ$ is straightforwardly extended to
arbitrary spinorial wave-functions by letting the operators act
component-wise on the $2s+1$ components. The canonical commutator
of $\wP$ and $\wQ$ \eqref{CCR} then carries over to spinor space
$\calH_{2}$ \eqref{Hs}. We can now appeal to the general
Lemma$\,$\ref{Lem1} and conclude that the two
arbitrary spin-particles are categorically and weakly discernible.
\tbf{Q.e.d.}\bskip
\indent Another possibility to finish the proof is more-or-less
to repeat the proof of Lemma$\,$\ref{Lem1} but now with
spinorial wave-functions. For step [S1], the case $N=2$,
the state space of the composite system becomes:
\beq \calH^{2}_{s}\,\equiv\,\big(L^{2}(\R^{3})\big)^{2s+1}\otimes
\big(L^{2}(\R^{3})\big)^{2s+1}\;, \label{H2s}
\enq
where the spinorial wave-functions now have $(2s+1)^{2}$
entries --- $(2s+1)^{N}$ for $N$ spin-$s$ particles.
A basis for $\calH^{2}_{s}$ \eqref{H2s} is
\beq \big\{\bfe_{k}\phi_{m}\otimes\bfe_{j}\phi_{l}\in\calH^{2}_{s}
\mid k,j\in\{1,2,\ldots,2s+1\},\;
m,l\in\N^{+}\big\}\;. \label{basisH2s} \enq
An arbitrary spinorial wave-function $\Psi\in\calH^{2}_{s}$ of
the composite system can then be expanded as follows:
\beq \Psi(\bfq_{1},\bfq_{2})\,=\,\sum_{k,j=1}^{2s+1}\,
\sum_{m,l=1}^{\infty}\frac{\gamma_{ml}}{(2s+1)}
\,\bfe_{k}\phi_{m}(\bfq_{1})\otimes\bfe_{j}\phi_{l}(\bfq_{2})\;,
\label{ExpPsi2} \enq
where the $\gamma_{ml}$ form a squarely-summable sequence,
i.e.\ a Hilbert-vector in $\ell^{2}(\N)$, of norm$\,1$.

With the usual definitions,
\beq \wP_{\pone}\,\equiv\,\wP\otimes\one,\quad
\wP_{\ptwo}\,\equiv\,\one\otimes\wP,\quad
\wQ_{\pone}\,\equiv\,\wQ\otimes\one,\quad
\wQ_{\ptwo}\,\equiv\,\one\otimes\wQ\;,
\enq
one obtains all the relevant commutators on $\calH^{2}_{s}$ by
using expansion \eqref{ExpPsi2}.
 The discerning relation \eqref{DefC}
on the direct-product spinor-space $\calH^{2}_{s}$ then becomes
\beq\sfC(\pa,\pb)\quadiff\all\Psi\in\calD:\;
\big[\wP_{\pa},\,\wQ_{\pb}\big]\Psi\,=\,-\rmi\hslash\Psi\;,
\label{CPQ}
\enq
where $\calD\subset\calH^{2}_{s}$ is the domain
of the commutator. Etc.

We close this Section with a number of systematic remarks.

\indent\emph{Remark$\:$1}. Notice that in contrast to the proof
of Theorem$\,$\ref{ThmDiscNPartInf}, the proof Corollary$\,$\ref{CorrSpin}
relies, besides on StateP, WkMP, WkPP and SemC, on the Symmetrisation Postulate (SymP)
\emph{only in so far as} that without this postulate the distinction between
integer and half-integer spin particles makes little sense and,
more importantly, the tacit claim that this distinction exhausts all
possible composite systems of similar particles is unfounded.
Besides this, SymP does not perform any deductive labour in the proof.
Specifically, the distinction between Bose-Einstein and
Fermi-Dirac states never enters the proof, which means that any restriction
on Hilbert-rays and on statistical operators,
as SymP demands, leaves the proof valid: the theorem holds
for all particles in all sorts of states,
fermions, bosons, quons, parons, quarticles, anyons and what have you.

\emph{Remark$\:$2}. The proofs of Theorem$\,$\ref{ThmDiscNPartInf}
and Corollary$\,$\ref{CorrSpin} exploit the non-commutativity of the
physical magnitudes, which is one of the algebraic
hall-marks of quantum physics.
Good thing. The physical meaning of relation $\sfC$ \eqref{CPQ}
can be understood as follows: momentum and position pertain to two
particles differently from how they pertain to 
a single particle. Admittedly this is something we already knew for
a long time, since the advent of $\qm$.
What we didn't know, but do know now, is that this
old knowledge provides the ground for discerning similar particles
weakly and categorically.

\emph{Remark$\:$3}. The spinorial wave-function $\Psi$ must
lie in the domain $\calD$ of the commutator of the unbounded operators
$\wP$ and $\wQ$, which domain
is a proper subspace of $\calH^{2}_{s}$ so that
the members of $\calH^{2}_{s}\!\setminus\!\calD$ fall outside
the scope of relation $\sfC$ \eqref{DefC}. The domain $\calD$
does however lies \emph{dense} in $\calH^{2}_{s}$, even the domain
of all polynomials of $\wP$ and $\wQ$ does so (the non-Abelian
ring on $\calD$ they generate) --- $\calD$ is
the Schwarz-space of all complex
wave-functions that are continuously differentiable
and fall off exponentially. This means that every wave-function that does
not lie in Schwarz-space can be approximated with
\emph{arbitrary} accuracy by means of wave-functions that do lie in Schwarz-space.
This is apparently good enough for physics. Then it is good enough for us too.

\emph{Remark$\:$4}. A special case of Theorem$\,$\ref{ThmDiscNPartInf}
is that two bosons in symmetric direct-product states, say
\beq \Psi(\bfq_{1},\bfq_{2})=\phi(\bfq_{1})\phi(\bfq_{2})\;,
 \label{DirectProdState} \enq
are also weakly discernible. This seems a hard nut to swallow.
If two bosons in state \eqref{DirectProdState} are \emph{discernible},
then something must have gone wrong. Perhaps we attach too much
metaphysical significance to a mathematical result?

Our position is the following. The weak discernibility of the
two bosons in state \eqref{DirectProdState} is a deductive consequence
of a few postulates of $\qm$. Rationality dictates that if one
accepts those postulates, one should accept every consequence of
those postulates. This is part of \emph{what it means to accept deductive
logic}, which we do accept. We admit that the discernibility
of two bosons in state \eqref{DirectProdState} is an unexpected
if not bizarre consequence. But in comparison to other bizarre consequences
of $\qm$, such as inexplicable correlations at a distance (\textsc{epr}),
animate beings that are neither dead nor alive (Schr\"{o}dinger's
immortal cat), kettles of water
on a seething fire that will never boil (quantum Zeno), an anthropocentric
and intentional concept taken as primitive (measurement), 
states of matter defying familiar
states of aggregation (\textsc{be}-condensate), in comparison to all that,
the weak discernibility of bosons in direct-product states is
not such a hard nut to swallow. Get real, it's peanuts.

\emph{Remark$\:$5}. Every `realistic' quantum-mechanical model of a physical
system, whether in atomic physics, nuclear physics or solid-state physics,
employs wave-functions. This means that \emph{now}, \emph{and only now},
we can conclude that the similar
elementary particles of the real world are categorically and weakly discernible.
Conjecture$\,1$ of Muller \& Saunders (2008: 537) has been proved.

Parenthetically, do finite-dimensional Hilbert-spaces
actually have applications at all?
Yes they have, in quantum optics and even more prominently in quantum information theory.
There one chooses to pay attention to spin-degrees of freedom only and ignores
all others --- position, linear momentum, energy. This is not
to deny there are physical magnitudes such as position, momentum or energy, or that
these physical magnitudes do not apply in the quantum-information-theoretic
models. Of course not. Ignoring these physical magnitudes
is a matter of expediency if one is not interested in them.
Idealisation and approximation are part and parcel of science.
No one would deny that quantum-mechanical models using spinorial
wave-functions in infinite-dimensional Hilbert-space match
physical reality \emph{better} --- if at all --- than
finite-dimensional models do that \emph{only} consider spin,
and it is for those better models that we have proved our case.

Nevertheless, we next proceed to prove the
discernibility of elementary particles for finite-dimensional
Hilbert-spaces.
\section{To Discern in Finite-Dimensional Hilbert-Spaces}%
\label{SectDiscThmsFin}
In the case of finite-dimensional Hilbert-spaces, considering
$\C^{d}$ suffices, because every $d$-dimensional Hilbert-space is
isomorphic to $\C^{d}$ ($d\in\N^{+}$). The proof is a vast
generalisation of the total-spin relation $\sfT$ of Muller \& Saunders
(2008: 535).
\begin{Thm}[\textbf{StateP, WkMP, StrPP, SymP}]~~In a composite physical system of
$N\geqslant 2$ similar particles, all particles are categorically weakly
discernible in every physical state, pure and mixed,
for every finite-dimensional Hilbert-space by only using their
spin degrees of freedom. \label{ThmDiscNPartFin}
\end{Thm}
\tbf{Proof.}
Let $\pa,\pb,\pj$ be particle-variables, ranging over the set
$\{\pone,\ptwo,\ldots,\pN\}$ of $N$ particles.
We proceed again Step-wise, as follows.\bskip
\indent [S1]~~Case for $N=2$, pure states.\newline
\indent [S2]~~Case for $N=2$, mixed states.\newline
\indent [S3]~~Case for $N>2$, all states.\bskip
[S1].~~\emph{Case for $N=2$, pure states.}
We begin by considering two similar particles,
labeled ${\pone,\ptwo}$, of spin-magnitude $s\hslash$,
which is a positive integer or a half-integer; $\pa$
and $\pb$ are again variables over this set. The single
particle Hilbert-space is $\C^{2s+1}$, which is isomorphic to
every $(s2+1)$-dimensional Hilbert-space; for $N$-particles
the associated Hilbert-space is the $N$-fold $\otimes$-product
of $\C^{2s+1}$. According to SymP, when we have considered
integer and half-integer spin particles, we have considered
all particles.

We begin by considering the spin-operator of a single
particle acting in $\C^{2s+1}$:
\beq \wbfS\,=\,\wS_{x}+\wS_{y}+\wS_{z}\;,
\enq
where $\wS_{x}$, $\wS_{y}$ and $\wS_{z}$ are the
three spin-operators along the three perpendicular spatial directions
($x,y,z$). The operators $\wbfS^{2}$ and $S_{z}$ are self-adjoint
and commute and therefore have a common set of orthonormal
eigenvectors $|s,m\ra$; their eigenvector-equations are:
\beq \wbfS^{\,2}|s,m\ra\,=\,s(s+1)\hslash^2|s,m\ra\quadand
\wS_{z}|s,m\ra\,=\,m\hslash\,|s,m\ra\;, \enq
where eigenvalue $m\in\{-s,-s+1,\ldots, s-1,+s\}$ (see e.g.\
Cohen-Tannoudji \emph{et al.}\ (1977: Ch.$\,$X) or
Sakurai (1995: Ch.$\,3$). Next we consider two particles.

The total spin operator of the composite system is
\beq
\wbfS\,\equiv\,\wbfS_{\pone}+\wbfS_{\ptwo}\;,\quad
\mrm{where}\quad\wbfS_{\pone}\,\equiv\,\wbfS\otimes\one\;,
\quad\wbfS_{\ptwo}\,\equiv\,\one\otimes\wbfS\;, \label{Stot}
\enq
and its $z$-component is
\beq \wS_{z}\,=\,
\wS_{z}\otimes\one+\one\otimes\wS_{z}\;, \enq
which all act in $\C^{2s+1}\otimes\C^{2s+1}$. The set
\beq \big\{\wbfS_{\pone},\;\wbfS_{\ptwo},\;\wbfS\;,
\;\wS_{z}\big\} \enq
is a set of commuting self-adjoint operators. These operators therefore have
a common set of orthonormal eigenvectors $|s;S,M\ra$.
Their eigenvector equations are:
\beq \begin{array}{ll}
\terug\wbfS_{\pone}^{\,2}|s;\,S,M\ra
&=\;s(s+1)\hslash^2\,|s;\,S,M\ra\;,
\meer\terug
\wbfS_{\ptwo}^{\,2}|s;\,S,M\ra
&=\;s(s+1)\hslash^2\,|s;\,S,M\ra\;,
\meer\terug
\wbfS^{\,2}|s;\,S,M\ra
&=\;S(S+1)\hslash^2\,|s;\,S,M\ra\;,
\meer\terug
\wS_{z}\,|s;\,S,M\ra
&=\;M\hslash\,|s;\,S,M\ra\;.
\end{array}\label{EqsSM}\enq
One easily shows that $S\in\{0,1,\ldots,2s\}$ and
$M\in\{-S,-S+1,\ldots,S-1,S\}$.

We note that every vector $|\phi\ra\in\C^{2s+1}\otimes\C^{2s+1}$
has a unique expansion in terms of these
orthonormal eigenvectors $|s;S,M\ra$,
because they span this space:
\beq |\phi\ra\,=\,\sum_{S=0}^{2s}\sum_{M=-S}^{+S}
\gamma(M,S)|s;\,S,M\ra\;, \label{Expphi}
\enq
where $\gamma(M,S)\in\C[0,1]$ and their moduli sum to~$1$.
Since the vectors $|s;m,m'\ra\equiv|s,m\ra\otimes|s,m'\ra$
also form a basis of $\C^{2s+1}\otimes\C^{2s+1}$, so that
\beq |\phi\ra\,=\,\sum_{m=-s}^{+s}\sum_{m'=-s}^{+s}
\alpha(m,m';\,s)|s;\,m,m'\ra\;, \label{Exp2phi}
\enq
where $\alpha(m,m';s)\in\C[0,1]$ and their moduli sum to~$1$,
these two bases can be expanded in each other. The expansion-coefficients
$\alpha(m,m';s)$ of the basis-vector $|s;S,M\ra$ are the
well-known `Clebsch-Gordon coefficients'. See for instance
Cohen-Tannoudji (1977: 1023).

Let us now proceed to prove Theorem \ref{ThmDiscNPartFin}.
Consider the following categorical `Total-spin relation':
\beq
\sfT(\pa,\pb)\quadiff\all|\phi\ra\in\C^{2s+1}\otimes\C^{2s+1}:\;
\big(\wbfS_{\pa}+\wbfS_{\pb}\big)^{2}\,|\phi\ra
\,=\,4s(s+1)\hslash^2\,|\phi\ra\;.\label{DefT}\enq
One easily verifies that relation $\sfT$ \eqref{DefT} meets
Req1 and Req2.

We now prove that relation $\sfT$ \eqref{DefT}
discerns the two fermions weakly.

\emph{Case$\,$1}: $\pa=\pb$. We then obtain the spin-magnitude operator of
a single particle, say $\pa$:
\beq
\big(\wbfS_{\pa}+\wbfS_{\pa}\big)^{2}\,|s;\,S,M\ra\,=\,
4\wbfS_{\pa}^{\,2}\,|s;\,S,M\ra\,
=\,4s(s+1)\hslash^2\,|s;\,S,M\ra\;,\label{Saa}\enq
which extends to arbitrary $|\phi\ra$ by expansion \eqref{Expphi}:
\beq \big(\wbfS_{\pa}+\wbfS_{\pa}\big)^{2}\,|\phi\ra\,=\,
4s(s+1)\hslash^2\,|\phi\ra\;.\label{Sphi}\enq

By WkPP, the composite system then possesses
the following quantitative physical property (when substituting $\pone$ or
$\ptwo$ for $\pa$):
\beq \big\la 4\wbfS_{\pa}^{2},\,4s(s+1)\hslash^2\big\ra\;.
\label{Aux3}\enq
This property \eqref{Aux3} \emph{is} a relation between the constituent
parts of the system, namely $\sfT$ \eqref{DefT}, and this relation is reflexive:
$\sfT(\pa,\pa)$ for every $\pa$ due to \eqref{Sphi}.

\emph{Case$\,$2}: $\pa\neq\pb$. The basis states $|s;\,S,M\ra$ are eigenstates
\eqref{EqsSM} of the total
spin-operator $\wbfS$ \eqref{Stot}:
\beq \big(\wbfS_{\pa}+\wbfS_{\pb}\big)^{2}\,|s;\,S,M\ra
\,=\,S(S+1)\hslash^2\,|s;\,S,M\ra\;, \enq
which does \emph{not} extend to arbitrary vectors $|\phi\ra$
but only to superpositions of basis-vectors having the
same value for $S$, that is, to vectors of the form:
\beq |s;\,S\ra\,=\,\sum_{M=-S}^{+S}
\gamma(M,S)|s;\,S,M\ra\;. \label{ExpS}
\enq
Since $S$ is maximally equal to $2s$, the eigenvalue
$S(S+1)$ belonging to vector $|s;S\ra$ \eqref{ExpS} is always smaller than $4s(s+1)=2s(s+1)+2s$,
because $s>0$. Therefore relation $\sfT$ \eqref{DefT}
fails for $\pa\neq\pb$ for all $S$:
\beq\big(\wbfS_{\pa}+\wbfS_{\pb}\big)^{2}\,|s;\,S,M\ra
\,\neq\,s(s+1)\hslash^2\,|s;\,S,M\ra\;. \enq
The composite system does indeed \emph{not} possess,
by SemC \eqref{SemC}, the following \emph{two} quantitative physical properties of
the composite system (substitute $\pone$ for $\pa$ and $\ptwo$ for $\pb$ or
conversely):
\beq \big\la\big(\wbfS_{\pa}+\wbfS_{\pb}\big)^{2},\;s(s+1)\hslash^2\big\ra\;,
\label{PropS} \enq
which is expressed by predicate $\sfT$ as a relation between its constituent parts,
$\pone$ and $\ptwo$, because the system does possess this property
according to WkPP:
\beq \big\la\big(\wbfS_{\pa}+\wbfS_{\pb}\big)^{2},\;S(S+1)\hslash^2\big\ra\;. \enq

However, superpositions of basis-vectors having a different
value for $S$, such as
\beq \mbox{\small $\frac{1}{\sqrt{2}}$}\big(|s;\,0,0\ra+|s;\,1,M\ra\big)\;,
\enq
where $M$ is $-1$, $0$ or $+1$, are \emph{not} eigenstates of
the total spin-operator \eqref{Stot}.
Precisely for these states we need to appeal to StrPP, because according to the converse of WkPP
this is sufficient to conclude that the composite system does not possess
physical property \eqref{PropS}, so that also for these states relation
$\sfT(\pa,\pb)$ fails for $\pa\neq\pb$. From this fact and the theorem of predicate logic \eqref{theoremlogic}, we then conclude that $\sfT$ is symmetric (Req2). Since
the operators involved correspond to
physical magnitudes, e.g.\ spin, relation $\sfT$ \eqref{DefT}
is physically meaningful (Req1) and hence is admissible, because
it also meets Req2 ($\sfT$ is reflexive and symmetric).

Therefore total-spin-relation $\sfT$ \eqref{DefT} discerns the
two spin-$s$ particles weakly. Since no probability
measures occur in the \emph{definiens} of $\sfT$; it discerns
them also categorically.

[S2].~~\emph{Case for $N=2$, mixed.}
The extension from pure to mixed states runs as before, as in
step [S2] of the proof of Lemma$\,$\ref{Lem1}.
There is however one subtle point we need to take care of.

\emph{Case$\,$1}: $\pa=\pb$.
Rewriting relation $\sfT$ \eqref{DefT} for $1$-dimensional projectors
is easy. Since the spin $s$ of the constituent particles is fixed,
the $1$-dimensional projector that projects on the ray that contains $|s;\,S,M\ra$
is an eigenoperator (eigenstate) of $(\wbfS_{\pa}+\wbfS_{\pa})^{2}$ having
the same eigenvalue $4s(s+1)$ \eqref{Saa}. Consequently, every (convex) sum of $1$-dimensional projectors
that project on vectors with the same value of $S$ has this same eigenvalue and we proceed
as before in [S2] of Lemma$\,$\ref{Lem1}, by an appeal to WkPP and a generalisation of
$\sfT$ \eqref{DefT} to mixed states:
\beq
\sfT(\pa,\pb)\quadiff\all W\in\calS(\C^{2s+1}\otimes\C^{2s+1}):\;
\big(\wbfS_{\pa}+\wbfS_{\pb}\big)^{2}\,W
\,=\,4s(s+1)\hslash^2\,W\;.\label{DefTW}\enq
For (convex) sums of projectors that project on vectors of different value of $S$,
we need StrPP again, as in step [S1] above. Relation $\sfT(\pa,\pa)$ \eqref{DefTW} holds also for mixed states.

\emph{Case$\,$2}: $\pa\neq\pb$.
The $1$-dimensional projector on $|s;S,M\ra$ now is an
eigenoperator (eigenstate) of $\big(\wbfS_{\pa}+\wbfS_{\pb}\big)^{2}$ having
eigenvalue $S(S+1)\hslash^{2}$ \eqref{EqsSM}. Since $S\leqslant 2s$ for every $S$,
this eigenvalue is necessarily smaller than $4s(s+1)$ for all $S$.
Then either every convex mixture of the $1$-dimensional projectors has 
 an eigenvalue smaller than $4s(s+1)$ too, or it is not an eigenstate of
$\big(\wbfS_{\pa}+\wbfS_{\pb}\big)^{2}$ at all (when the mixture consists of
projectors on different states $|s;S,M\ra$ and $|s;S',M'\ra$, $S\neq S'$).
In virtue of StrPP, relation $\sfT$ \eqref{DefT} then does \emph{not} hold for
its parts (for $\pa\neq\pb$), for \emph{all} states, mixed and pure, because the
system does not possess the required physical property.

So $\sfT$ \eqref{DefTW} is reflexive and symmetric (Req2) and certainly
physically meaningful (Req1). In conclusion two similar particles in in every
finite-dimensional are categorically weakly discernible in all admissible states,
both pure and mixed.

[S3].~~\emph{Case for $N>2$, all states.}
Consider a subsystem of two particles, say $\pa$ and $\pb$, of
the $N$-particle system.
We can consider these two to form a composite system and then repeat
the proof we have just given, in [S1] and [S2], to show they are weakly and categorically
discernible. When we can discern an arbitrary particle, say $\pa$, from
every other particle, we have discerned all particles.
\tbf{Q.e.d.}\newline\newline
\noindent We end this Section again with a few more systematic remarks.

\emph{Remark$\:$1}. In our proofs we started with $N$
particles. Is it not circular, then, to prove they are discernible because
to \emph{assume} they are \emph{not identical} (for if they were, we would
have single particle, and not $N>1$ particles), implies we are somehow tacitly \emph{assuming} they are discernible? Have we committed
the fallacy of propounding a \emph{petitio principii}?

No we have not. We assume the particles are \emph{formally} discernible,
e.g.\ by their labels, but then
demonstrate on the basis of a few postulates of $\qm$ that they
are \emph{physically} discernible. Or in other words,
we assume the particles are quantitatively not-identical and
we prove they are qualitatively not-identical. Or still in other
words, we assume numerical diversity and prove weak qualitative
diversity. See further Muller \& Saunders (2008:
541--543) for an elaborate discussion of precisely this issue.

\emph{Remark$\:$2}. Of course Theorem$\,$\ref{ThmDiscNPartInf}
implies probabilistic versions.
The \emph{Probability Postulate} (ProbP) of $\qm$ gives the Born probability measure over measurement outcomes for pure states and gives Von Neumann's extension to mixed states,
which is the trace-formula. By following the strategy of Muller \& Saunders
(2008: 536--537) to carry over categorical proofs
to probabilistic proofs, one easily proves the probabilistic
weak discernibility of similar particles, notably then without
using WkPP and SemC \eqref{SemC}.

\emph{Remark$\:$3}. In contrast to Theorem$\,$\ref{ThmDiscNPartInf},
Theorem$\,$\ref{ThmDiscNPartFin} relies on StrPP, which arguably
is an empirically superfluous postulate. StrPP also leads almost
unavoidably to nothing less than the Projection Postulate
(see Muller \& Saunders 2008: 514). Foes of the Projection Postulate
are not committed to Theorem$\,$\ref{ThmDiscNPartFin}.
They will find themselves metaphysically in the following situation
(provided they accept the whiff of interpretation WkPP):
similar elementary particles in infinite-dimensional
Hilbert-spaces are weakly discernible, in certain classes of states in
finite-dimensional Hilbert-spaces they are also weakly discernible,
fermions in finite-dimensional Hilbert-spaces are weakly discernible
in all admissible states when there always is a maximal operator
of physical significance (see Introduction),
but for other classes of states in finite-dimensional Hilbert-spaces the
jury is still out.

For those who have no objections against StrPP, all similar particles in all
kinds of Hilbert-spaces in all kinds of states are weakly discernible.
This may be seen as an argument in favour of StrPP: it leads
to a uniform nature of elementary particles when
described quantum-mechanically and the proofs make
no distinction between fermions and bosons.

\emph{Remark$\:$4}. The so-called \emph{Second Underdetermination Thesis}
says roughly that the physics underdetermines the metaphysics ---
the \emph{First Underdetermination Thesis} then is the familiar Duhem-Quine
thesis of the underdetermination of theory by all actual or by all
possible data; see Muller (2009). \emph{Naturalistic metaphysics}, as
recently has been vigorously defended by Ladyman \& Ross (2007: 1--65),
surely follows scientific theory wherever scientific theory leads us,
without prejudice, without clinging to so-called common sense,
and without tacit adherence to
what they call \emph{domesticated metaphysics}. Well, $\qm$ leads us
by means of mathematical proof to the metaphysical statements
(if they are metaphysical) that similar elementary particles are
categorical (and by implication probabilistic) relationals,
more specifically weak discernibles.
Those who have held that $\qm$ underdetermines the metaphysics
in this regard (see references in the Introduction),
in this case the nature of the elementary particle, are
guilty of engaging in unnatural metaphysics (for elaboration, see Muller (2009: Section$\,4$)).
\section{Conclusion: Leibniz Reigns}%
\label{SectConclusion}%
We have demonstrated that for every set
$\calS_{N}$ of $N$ similar particles, in infinite-dimensional and
finite-dimension Hilbert-spaces, in all their physical states, pure and mixed,
similar particles can be discerned by
physically meaningful and permutation-invariant means,
and therefore are \emph{not} physically indiscernible:
\beq
\QM^{-}\;\;\entails\;\;\all N\in\{2,3,\ldots\},\;
\all\pa,\pb\in\calS_{N}:\;\pa\neq\pb\;\ifthen\;
\deny\msf{PhysInd}(\pa,\pb)\;, \enq
where $\QM^{-}$ now stands for StP, WkMP, StrPP and SymP,
which is logically the same as having proved $\pii$ \eqref{PIIQM}:
\beq
\QM^{-}\;\;\entails\;\;\all N\in\{2,3,\ldots\},\;
\all\pa,\pb\in\calS_{N}:\;\msf{PhysInd}(\pa,\pb)\;\ifthen\;\pa=\pb\;,
\label{ThmPII}\enq
and by theorem of logic \eqref{PIIIT} as having disproved $\it$. Hence
\beq
\QM^{-}\;\;\entails\;\;\,\pii\;\conj\;\deny\it\;.
\label{EndResult}\enq
Therefore all claims to the contrary, that $\qm$ refutes $\pii$,
or is inconsistent with $\pii$, or that $\pii$ cannot be established (see
the references in Section$\,$\ref{SectIntro} for propounders of these
claims) find themselves in heavy weather.
Quantum-mechanical particles are \emph{categorical weak discernibles}, and
therefore not \emph{indiscernibles} as propounders of $\it$ have claimed.
Similar elementary particles are like points on a line, in a plane or
in Euclidean space: absolutely indiscernible yet not identical
(there is more than one of them!). Points on a line are categorical
relationals, categorical weak discernibles to be precise.
Elementary particles are exactly like points in this regard.

Leibniz is back from exile and reigns over all
quantum-mechanically possible worlds, \emph{salva veritate}.
\clearpage\section*{References}%
\addcontentsline{toc}{section}{References}%
{\small
Barnette, R.L. (1978), `Does Quantum Mechanics Disprove the
Principle of the Identity of Indiscernibles?', \emph{Philosophy of
Science} \tbf{45} (1978) 466--470.\bskip
Brading, K., Castellani, E. (2003), \emph{Symmetries in
Physics: New Reflections}, Cambridge: Cambridge University Press,
2003.\bskip
Butterfield, J.N. (1993), `Interpretation and Identity in Quantum
Theory', \emph{Studies in History and Philosophy of Science}
\tbf{24} (1993) 443--476.\bskip
Castellani, E. (1998), \emph{Interpreting Bodies.\ Classical and
Quantum Objects in Modern Physics}, Princeton, New Jersey:
Princeton University Press, 1998.\bskip
Castellani, E., Mittelstaedt, P. (1998), `Leibniz's Principle,
Physics and the Language of Physics', \emph{Foundations of
Physics}, \tbf{30} (2000) 1587--1604.\bskip
Cohen-Tannoudji, C. et al. (1977),
\emph{Quantum Mechanics}, Volume$\,$II, New York: John Wiley \&
Sons, 1977.\bskip
Cortes, A. (1976), `Leibniz's Principle of the Identity of
Indiscernibles: A False Principle', \emph{Philosophy of Science}
\tbf{43} (1976) 491--505.\bskip
Dalla Chiara, M.L., Toraldo di Francia, G. (1993),
`Individuals, Kinds and Names in Physics', in Corsi, G., et al.
(eds.), \emph{Bridging the Gap: Philosophy, Mathematics, Physics},
Dordrecht: Kluwer Academic Publishers, pp.~261--283.\bskip
Dalla Chiara, M.L., Giuntini, R., Krause, D. (1998), `Quasiset
Theories for Micro-objects: A Comparision', in Castellani (1998,
142--152).\bskip
Fraassen, B.C.\ van. (1991),
\emph{Quantum Mechanics.\ An Empiricist View}, Oxford: Clarendon
Press, 1991.\bskip
French, S., Redhead, M.L.G. (1988),
`Quantum Physics and the Identity of Indiscernibles, \emph{British
Journal for the Philosophy of Science} \tbf{39} (1988) 233--246.
\bskip
French, S. (1989a), `Why the Principle of the Identity of
Indiscernibles is not contingently True Either', \emph{Synthese}
\tbf{78} (1989) 141--166.\bskip
French, S. (1989b), `Identity and Indiscernibility in Classical
and Quantum Physics', \emph{Australasian Journal of Philosophy}
\tbf{67} (1989) 432--446.\bskip
French, S. (1998), `On the Withering Away of Physical Objects', in
Castellani (1998, 93--113).\bskip
French, S., Rickles, D. (2003), `Understanding Permutation
Symmetry', in Brading \& Castellani (2003, 212--238).\bskip
French, S. (2006), `Identity and Individuality in Quantum Theory',
\emph{Stanford Encyclopedia of Philosophy}, E.N.\ Zalta (ed.),
\newline \textsc{url} = $\la$
http://plato.stanford.edu/archives/spr2006/entries/qt-idind/$\ra$
\bskip
French, S., Krause, D. (2006), \emph{Identity in Physics: A
Historical, Philosophical and Formal Analysis}, Oxford: Clarendon
Press, 2006.\bskip
Huggett, N. (2003), `Quarticles and the Identity of
Indiscernibles', in Brading \& Castellani (2003, pp.~239--249).
\bskip
Krause, D. (1992), `On a Quasi-set Theory', \emph{Notre Dame
Journal of Formal Logic} \tbf{33} (1992) 402--411.\bskip
Ladyman, J. (2007), `On the Identity and Diversity of Objects in a Structure',
\emph{Proceedings of the Aristotelian Society}, Supplementary Volume \tbf{81}
(2007) 23--43.\bskip
Ladyman, J., Ross, D. (2007), \emph{Every Thing Must Go.\ Metaphysics
Naturalized}, Oxford: Oxford University Press, 2007.\bskip
Margenau, H. (1944), `The Exclusion Principle and its
Philosophical Importance', \emph{Philosophy of Science} \tbf{11}
(1944) 187--208.\bskip
Massimi, M. (2001), `The Exclusion Principle and the Identity of
Indiscernibles: a Response to Margenau's Argument', \emph{British
Journal for the Philosophy of Science} \tbf{52} (2001)
303--330.\bskip
Muller, F.A. (2009), `Withering Away, Weakly', to appear in
\emph{Synthese}.\bskip
Muller, F.A., Saunders, S.W. (2008), `Discerning Fermions', \emph{British
Journal for the Philosophy of Science} \tbf{59} (2008) 499--548.\bskip
Redhead, M.L.G., Teller, P. (1992), `Quantum Physics and the
Identity of Indiscernibles', \emph{British Journal for the
Philosophy of Science} \tbf{43} (1992) 201--218.\bskip
Sakurai, J.J. (1995). \emph{Modern Quantum Mechanics},
Revised Edition, New York: Addison Wesley Publishing Company.\bskip
Saunders, S. (2006), `Are quantum particles objects?',
\emph{Analysis} \tbf{66.1} (2006) 52--63.\bskip
Teller, P. (1998), `Quantum Mechanics and Haecceities', in
Castellani (1991, 114--141).\bskip
Weyl, H. (1931), \emph{The Theory of Groups and Quantum
Mechanics}, London: Methuen \& Company, 1931.\bskip
}
\end{document}